\documentclass[12pt]{article}
\usepackage[cp1251]{inputenc}
\usepackage{amsfonts}
\usepackage{latexsym}
\def\beq{\begin{equation}}
\def\eeq{\end{equation}}
\def\bea{\begin{eqnarray}}
\def\eea{\end{eqnarray}}
\begin{document}

\rightline{SINP MSU 2005-30/796}
\begin{center}
{\Large \bf Physical degrees of freedom \\
  in stabilized brane world models  } \\

\vspace{4mm}

Edward E. Boos, Yuri S. Mikhailov$^a$,  Mikhail N. Smolyakov, Igor
P.~Volobuev\\ \vspace{0.5cm} Skobeltsyn Institute of Nuclear
Physics, Moscow State University \\ 119992 Moscow, Russia \\ $^a$
Physics Department, Moscow State University \\ 119992 Moscow,
Russia \\
\end{center}

\begin{abstract}
We consider brane world models with interbrane separation
stabilized by the Goldberger-Wise scalar field. For arbitrary
background, or vacuum configurations of  the gravitational and
scalar fields in such models, we construct the second variation
Lagrangian, study its gauge invariance, find the corresponding
equations of motion and decouple them in a suitable gauge. We also
derive an effective four-dimensional Lagrangian for such models,
which describes the massless graviton, a tower of massive
gravitons and a tower of massive scalars. It is shown that for a
special choice of the background solution the masses of the
graviton excitations may be of the order of a few TeV, the radion
mass of the order of 100 GeV, the inverse size of the extra
dimension being tens of GeV. In this case the coupling of the
radion to matter on the negative tension brane is approximately
the same as in the unstabilized model with the same values of the
fundamental five-dimensional energy scale and the interbrane
distance.
\end{abstract}

\section{Introduction}
Brane world models and their phenomenology have been widely
discussed in the last years \cite{1}--\cite{Rubakov:2001kp}. One
of the most interesting brane world models is the Randall-Sundrum
model with two branes, - the RS1 model \cite{Randall:1999ee}. This
model solves the hierarchy problem due to the warp factor in the
metric and predicts an interesting  new physics in the TeV range
of energies. A flaw of the RS1 model  is the presence of a
massless scalar mode, called the radion, which arises due to the
fluctuations of the branes with respect to each other. Its
interactions contradict the existing experimental data, and in
order the model be phenomenologically acceptable the radion must
acquire a mass, which is equivalent to the stabilization of the
brane separation distance. The latter can be achieved by
introducing a five-dimensional scalar field with bulk and brane
potentials, whose vacuum energy has a minimum for a certain
interbrane distance \cite{wise}. A disadvantage of the approach
proposed in \cite{wise} is that the backreaction of the scalar
field on the background metric is not taken into account. This
problem is solved in the model proposed in \cite{wolfe}.

Nevertheless, most of the papers on the phenomenology of the RS1
model consider the unstabilized model, just putting the radion
mass by hand. Such an approach seems to be inconsistent, because
the backreaction of the scalar field on the metric leads to a
renormalization of the parameters of the RS1 model. The scalar
sector of the stabilized RS1 model in the background of
\cite{wolfe} was studied in \cite{Csaki:2000zn},
\cite{Kofman:2004tk}, where the fundamental five-dimensional
energy scale of the theory was assumed to be of the order of
four-dimensional Planck mass. In \cite{Boos:2004uc} it was shown
that a consistent physical interpretation of the theory on the
negative tension brane for the model of \cite{wolfe} is possible,
if the fundamental five-dimensional scale is of the order of $1
TeV$, rather than the Planck one.

In the present paper we are going to study all the physical
degrees of  freedom of the stabilized RS1 model, i.e. both the
tensor and the scalar sectors,  in an arbitrary background. Our
approach is based on the Lagrangian description of the linearized
gravity, which was developed for the unstabilized model in
\cite{Boos:2002ik}. We find a convenient gauge and decouple the
equations for tensor and scalar modes. Then for the background
solution of \cite{wolfe} we calculate the masses of tensor and
scalar excitations and their couplings to matter on the negative
tension brane.

\section{Linearized gravity in stabilized brane world models}
Let us denote the coordinates  in five-dimensional space-time
$E=M_4\times S^{1}/Z_{2}$  by $\{ x^M\} \equiv \{x^{\mu},y\}$, $M=
0,1,2,3,4, \, \mu=0,1,2,3 $, the coordinate $x^4 \equiv y, \quad
-L\leq y \leq L$ parametrizing the fifth dimension. It forms the
orbifold, which is realized as the circle of circumference $2L$
with the points $y$ and $-y$ identified. Correspondingly, the
metric $g_{MN}$  and the scalar field $\phi$ satisfy the orbifold
symmetry conditions
\begin{eqnarray}
\label{orbifoldsym}\nonumber
 g_{\mu \nu}(x,- y)=  g_{\mu \nu}(x,  y), \quad
  g_{\mu 4}(x,- y)= - g_{\mu 4}(x,  y), \\
   g_{44}(x,- y)=  g_{44}(x,  y), \quad
   \phi(x,- y)=  \phi(x,  y).
\end{eqnarray}
The branes are located at the fixed points of the orbifold, $y=0$
and $y=L$.

The  action of stabilized brane world  models can be written as
\begin{equation}\label{actionDW}
S = S_g + S_\phi,
\end{equation}
where $S_g$ and $S_\phi$ are given by
\begin{eqnarray}\nonumber
S_g= 2 M^3\int d^{4}x \int_{-L}^L dy  R\sqrt{-g},\\ \nonumber
S_\phi = -\int d^{4}x \int_{-L}^L dy \left(\frac{1}{2}
g^{MN}\partial_M\phi\partial_N\phi+V(\phi)\right)\sqrt{-g} -\\
\label{actionsDW}  -\int_{y=0} \sqrt{-\tilde g}\lambda_1(\phi)
d^{4}x -\int_{y=L}\sqrt{-\tilde g}\lambda_2(\phi)    d^{4}x .
\end{eqnarray}
Here $V(\phi)$ is a bulk scalar field potential and
$\lambda_{1,2}(\phi)$ are brane   scalar field potentials,
$\tilde{g}=det\tilde g_{\mu\nu}$, and $\tilde g_{\mu\nu}$  denotes
the metric induced on the branes. The signature of the metric
$g_{MN}$ is chosen to be $(-,+,+,+,+)$.

The standard ansatz  for the  metric and the scalar field, which
preserves the Poincar\'e invariance in any four-dimensional
subspace $y=const$, looks like
\begin{eqnarray}\label{metricDW}
&ds^2=  e^{-2A(y)}\eta_{\mu\nu}  {dx^\mu  dx^\nu} +  dy^2 \equiv
\gamma_{MN}(y)dx^M dx^N,& \\ &\phi(x,  y) = \phi(y),&
\end{eqnarray}
 $\eta_{\mu\nu}$ denoting the flat Minkowski metric. If one substitutes
 this ansatz into the equations corresponding to action
(\ref{actionDW}), one gets a rather complicated system of
nonlinear differential equations for functions $A(y),\phi(y)$:
\begin{eqnarray}\nonumber
\frac{d V}{d\phi}+\frac{d\lambda_1 }{d\phi}\delta(y)
+\frac{d\lambda_2 }{d\phi}\delta(y-L)= -4A'\phi'+\phi'',\\
\nonumber 12M^3 (A')^2+\frac{1}{2}(V-\frac{1}{2} (\phi')^2)=0, \\
\label{yd}
\frac{1}{2}\left(\frac{1}{2}(\phi')^2+V+\lambda_1\delta(y)+\lambda_2\delta(y-L)
\right)=-2M^3\left(-3A''+6(A')^2\right).
 \end{eqnarray}
Here $'=\partial_4\equiv{\partial}/{\partial{y}}$.

Suppose we have a solution $A(y), \phi(y)$ to this system for an
appropriate choice of the parameters of the potentials such that
the interbrane distance is stabilized and is equal to $L$.  It
means that the vacuum energy of the scalar field has a minimum for
this value of the interbrane distance.

Now the linearized theory is obtained by representing the metric
and the scalar field as
\begin{eqnarray}\label{metricparDW}
g_{MN}(x,y)&=& \gamma_{MN}(y) + \frac{1}{\sqrt{2M^3}} h_{MN}(x,y),
\\ \label{metricparDW1}
\phi(x,y) &=& \phi(y) + \frac{1}{\sqrt{2M^3}} f(x,y),
\end{eqnarray}
substituting this representation into action (\ref{actionDW}) and
keeping the terms of the second order  in $h_{MN}$ and $f$. The
Lagrangian of this action is called the second variation
Lagrangian and has the form
\begin{eqnarray}\nonumber
&\frac{\mathcal L}{\sqrt{-\gamma}}
=-\frac{1}{4}(\bigtriangledown_S h_{MN} \bigtriangledown^S
h^{MN}+2\bigtriangledown_N h \bigtriangledown_M h^{MN}-& \\
\nonumber &- 2\bigtriangledown_M h^{MN}\bigtriangledown^S
h_{SN}-\bigtriangledown_S h \bigtriangledown^S h )+(A')^2 \left(
\frac {7}{2} h_{MN}h^{MN}-hh\right)-\\  \nonumber
&-A''\left(h_{MN}h^{MN}-\frac{1}{2}
\tilde{h}h+\frac{1}{2}h_{M\nu}h^{M\nu}\right)+\frac{1}{4M^3}
\biggl[ \frac{V}{2}\left(h_{MN}h^{MN}-\frac{1}{2}hh \right)+& \\
\nonumber &+\frac{1}{2} \left( h_{\mu\nu}
h^{\mu\nu}-\frac{1}{2}\tilde{h}\tilde{h}
\right)[\lambda_1\delta(y)+\lambda_2\delta(y-L)]+& \\ \nonumber
&+\frac{1}{2}(\phi')^2
\biggl(-\frac{1}{4}hh+\frac{1}{2}h_{MN}h^{MN}+hh_{44}-2h_{4M}h^{4M}
\biggr)-& \\ \nonumber &-f\biggl(h\frac{d
V}{d\phi}+\tilde{h}\bigl[
\frac{d\lambda_1}{d\phi}\delta(y)+\frac{d\lambda_2}
{d\phi}\delta(y-L)\bigr] \biggr)-f'\phi'h+2\partial_M f\phi'
h^{M4}-& \\ \label{l} &-\partial^Mf\partial_Mf-f^2\biggl(\frac{d^2
V }{d \phi^2}+\frac{d^2 \lambda_1 }{d \phi^2}\delta(y)+\frac{d^2
\lambda_2 }{d \phi^2}\delta(y-L) \biggr) \biggr]. &
\end{eqnarray}
Here $\; h=\gamma_{MN}h^{MN}, \;
\tilde{h}=\gamma_{\mu\nu}h^{\mu\nu}, \;$ $\phi$ stands for the
background solution and  $\bigtriangledown_M$ denotes the
covariant derivative with respect to metric $\gamma_{MN}$.

Varying the action built with this Lagrangian and taking into
account background field equations (\ref{yd}), we arrive at the
following equations of motion for the fluctuations of metric:
\begin{enumerate}\item $\mu\nu$-component
\begin{eqnarray}\nonumber
&\frac{1}{2}(\partial_\sigma{\partial^\sigma{h_{\mu\nu}}}-\partial_\mu
{\partial^\sigma{h_{\sigma\nu}}}-\partial_\nu{\partial^\sigma{h_{\sigma\mu}}}
+\partial_4{\partial_4{h_{\mu\nu}}})+\frac{1}{2}\partial_\mu{\partial_
\nu{\tilde{h}}}+&\\ \nonumber
&+\frac{1}{2}\partial_\mu{\partial_\nu{h_{44}}}-\frac{1}{2}\partial_4{
(\partial_\mu{h_{4\nu}}+\partial_\nu{h_{4\mu}})}
+A'(\partial_\mu{h_{4\nu}}+
\partial_\nu{h_{4\mu}})+& \\ \nonumber &+\frac{1}{2}\gamma_{\mu\nu}\biggl(- \partial_4 {\partial_4
{\tilde{h}}} -  \partial_{\sigma}{\partial^{\sigma}
{h_{44}}}-\partial_\sigma{\partial^\sigma
{\tilde{h}}}+4A'\partial_4{\tilde{h}}-3A'\partial_4{h_{44}} +&\\
\nonumber &+\partial_{\sigma}{\partial_{\tau}{h^{\sigma\tau}}}+
2\partial^{\sigma}{\partial_4{h_{\sigma 4}}}-
4A'\partial^\sigma{h_{4 \sigma}} \biggr)-h_{\mu\nu}(2A'^2-A'')+&\\
\label{munu}
&+\frac{3}{2}h_{44}\gamma_{\mu\nu}(4A'^2-A'')-\frac{1}{4M^3}\gamma_{\mu\nu}\left(f'\phi'+
f(-4A'\phi'+ \phi'') \right)=0; &
\end{eqnarray}
\item $\mu 4$-component
\begin{eqnarray}\nonumber
&\partial_4(\partial_\mu\tilde{h}-\partial^{\nu}
h_{\mu\nu})+\partial^\nu(\partial_\nu h_{\mu 4 }-\partial_\mu
h_{\nu 4})+3A'\partial_{\mu} h_{44}+&\\ \label{hm4} &+
\frac{1}{2M^3}\partial_\mu f\phi'=0;&
\end{eqnarray}
\item $44$-component
\begin{eqnarray}\nonumber
&\partial^\mu(\partial^\nu h_{\mu\nu}-\partial_{\mu}
\tilde{h})-6A'\partial^{\mu} h_{\mu4}
+3A'\partial_4{\tilde{h}}+&\\ \label{44eq} &+
\frac{1}{2M^3}\left(- h_{44}V-f\frac{d
V}{d\phi}+f'\phi'\right)=0;&
\end{eqnarray}
\item equation for the field $f$
\begin{eqnarray}\nonumber
&h_{44}\left( \frac{d\lambda_1}{d\phi}\delta(y)+\frac{d\lambda_2}
{d\phi}\delta(y-L)\right)-2h_{44}(-4A'\phi'+\phi'')+&\\ \nonumber
&+\phi'\partial_4\tilde{h}-\phi'\partial_4
h_{44}-2\phi'\partial^{\mu}h_{\mu4}-8A'\partial_4f
+2\partial_M{\partial^M{f}}-&\\ \label{eqf} &-2f\left(\frac{d^2 V
}{d \phi^2}+\frac{d^2 \lambda_1 }{d \phi^2}\delta(y)+\frac{d^2
\lambda_2 }{d \phi^2}\delta(y-L) \right)=0.&
\end{eqnarray}
\end{enumerate}
We will also use the following auxiliary equation, which is
obtained by contracting the indices in the $\mu\nu$-equation:
\begin{eqnarray}\label{cont}\nonumber
&\partial^\mu \partial^\nu h_{\mu\nu}- \partial^\mu \partial_\mu
(\tilde{h}+\frac{3}{2}h_{44})-
6A'\partial_4(h_{44}-\tilde{h})-\frac{3}{2}\partial_4
\partial_4\tilde{h}-6A'\partial^\mu h_{\mu 4}+&\\
&+3\partial^\mu
\partial_4 h_{\mu 4}+6h_{44}(4A'^2-A'')
-{\frac{1}{M^3}}(f'\phi'+f(-4A'\phi'+\phi''))=0.&
\end{eqnarray}
These equations were also discussed in \cite{Boos:2004uc}. The
normalization of fields in  the above  equations, as well as in
Eqs. (\ref{metricparDW}), (\ref{metricparDW1}) and (\ref{l})
differs from the one adopted in this paper.  It is more convenient
in the Lagrangian approach and does not effect the equations of
motion.

These equations are invariant under the gauge transformations
\begin{eqnarray}\nonumber
&h_{\mu\nu}^{(\prime)}(x,y)=h_{\mu\nu}(x,y)-(\partial_\mu{\xi_\nu}+\partial_\nu{
\xi_\mu}-2\gamma_{\mu\nu}\partial_4{A}\xi_4),&\\ \nonumber &
h_{\mu4}^{(\prime)}(x,y)=h_{\mu4}(x,y)-(\partial_\mu{\xi_4}+\partial_4{\xi_
\mu}+2\partial_4{A}\xi_\mu),&\\ \label{kal}
&h_{44}^{(\prime)}(x,y)=h_{44}(x,y)-2\partial_4{\xi_4},&\\\label{kalf}
&f^{(\prime)}(x,y)=f(x,y)-\partial_4\phi\xi_4,
\end{eqnarray}
provided ${\xi_M(x,y)}$ satisfy the orbifold symmetry conditions
$$ \xi_{\mu}(x,-y)= \xi_{\mu}(x,y), \quad \xi_4(x,-y)=
-\xi_4(x,y). $$ These gauge transformations are a generalization
of the gauge transformations in the unstabilized RS1 model
\cite{Rubakov:2001kp,Boos:2002ik}. We will use them to isolate the
physical degrees of freedom of the fields $h_{MN}$ and $f$. Let us
show that the gauge transformations with function $\xi_4$ allow
one to impose the gauge condition
\begin{equation}\label{uf}(e^{-2A}h_{44})'-\frac{1}{3
M^3}e^{-2A}\phi'f=0.
\end{equation}
Really, Eqs. (\ref{kal}), (\ref{kalf}) imply the following
equation for $\xi_4$ $$\partial_4{\partial_4{\xi_4}} -
2A'\partial_4{\xi_4}-\frac{1}{6M^3}(\phi')^2
\xi_4=-\frac{1}{2}(\partial_4{h_{44}} -
2A'h_{44}-\frac{1}{3M^3}\phi'f).$$ The standard theory of
differential equations demands the coefficient in front of
$\partial_{4}\xi_4$ to be continuous, though in our case it is not
so. To cure this flaw, we rewrite the equation as follows:
\begin{eqnarray} \nonumber
\partial_4\partial_4(\xi_4 e^{-A})+\xi_4
e^{-A}\left(A''-(A')^2-\frac{1}{6M^3}(\phi')^2\right)=
\\ \label{ksi}  =-\frac{1}{2}e^{-A}(\partial_4{h_{44}} -
2A'h_{44}-\frac{1}{3M^3}\phi'f)\equiv w(x,y).
\end{eqnarray}
The functions  $(A')^2$ and $(\phi')^2$ are smooth functions of
$y$. Although function  $A^{\prime \prime}$ has $\delta$-like
singularities at $y=0, y=L$, the singular terms drop from the
equation, because $\xi_4$ is equal to zero at these points. Thus,
the factor in front of $\xi_4 e^{-A}$ is a smooth function, and
the equation for $\xi_4 e^{-A}$ in the interval $[-L,L]$ can be
treated by the standard methods. In accordance with the general
theory, the homogeneous equation, corresponding to (\ref{ksi}),
has two independent solutions in the interval $[0,L]$; we denote
them $\chi_1(y)$ and $\chi_2(y)$. Now we can use the Green
function method to find $\xi_4$, which gives:
\begin{eqnarray}\nonumber
\xi_4(x,y)=e^A\int_0^L\frac{\chi_1(y)\chi_2(z)-\chi_2(y)
\chi_1(z)}{W(\chi_1(z)\chi_2(z))}\theta(y-z)w(x,z)dz-\\
-e^A\frac{\chi_1(y)\chi_2(0)-\chi_2(y)\chi_1(0)}{\chi_1(L)\chi_2(0)-\chi_2(L)
\chi_1(0)}\int_0^L\frac{\chi_1(L)\chi_2(z)-\chi_2(L)\chi_1(z)}
{W(\chi_1(z)\chi_2(z))}w(x,z)dz.
\end{eqnarray}
Here $W(\chi_1(z)\chi_2(z))$ stands for the Wronskian of the
solutions  $\chi_1(y)$ and $\chi_2(y)$, which in the case of Eq.
(\ref{ksi})  is just  a constant. It is not difficult to check
that the obtained function $\xi_4$ is equal to zero at the ends of
the interval and therefore can be continued to an odd function on
$[-L,L]$. Thus, gauge condition (\ref{uf}) really exists. We also
note that this relation was obtained in \cite{Csaki:2000zn} from
the equation for $\mu4$-component, in which only the scalar
degrees of freedom were retained. Similar to the case of the
unstabilized RS1 model, the gauge transformations with functions
$\xi_\mu$ allow one to impose the gauge $h_{\mu4}(x,y)=0$, after
which there remain the gauge transformations satisfying
\begin{equation}\label{restr}\partial_4({e^{2A}\xi_\mu})=0.
\end{equation}
Thus, we can use the gauge
\begin{eqnarray}\nonumber
(e^{-2A}h_{44})'-\frac{1}{3 M^3}e^{-2A}\phi'f=0, \\ \label{gauge}
h_{\mu 4} =0.
\end{eqnarray}

Next we  represent the gravitational field as
\begin{equation}\label{hb}
h_{\mu\nu}=b_{\mu\nu}+\frac{1}{4}\gamma_{\mu\nu}\tilde{h},
\end{equation}
with  $b_{\mu\nu}$ being a traceless tensor field
($\gamma^{\mu\nu}b_{\mu\nu}=0$).

Substituting gauge conditions (\ref{gauge}) and representation
(\ref{hb}) into the $\mu4$-equation and into  contracted
$\mu\nu$-equation  (\ref{cont}), we get:
\begin{eqnarray} \label{constraintg}
-\partial_4 (\partial^\nu b_{\mu\nu} ) + \frac{3}{4}\partial_\mu
\partial_4(\tilde h + 2 h_{44}) =0, \\ \nonumber
\partial^\mu \partial^\nu b_{\mu\nu} -\frac{3}{4} \partial_\rho
\partial^\rho \tilde h-\frac{3}{2} \partial_\rho \partial^\rho
h_{44} -\frac{3}{2} \frac{\partial^2} { \partial y^2} \tilde h +\\
\label{contg} + 6A' \partial_4\tilde h -3\frac{\partial^2}{
\partial y^2} h_{44} + 12 A'\partial_4 h_{44}=0.
\end{eqnarray}

Eq. (\ref{constraintg}) suggest the substitution $\tilde h = - 2
h_{44}$, which allows one to decouple the equations for the fields
$b_{\mu\nu}, h_{44}$ and  $f$. The possibility of using this
substitution and representation (\ref{hb}) for decoupling tensor
and scalar equations was mentioned in \cite{Kofman:2004tk}.

Really, as a result of this substitution Eqs. (\ref{constraintg}),
(\ref{contg}) take the form
\begin{eqnarray}\label{db}
\partial_4 (\partial^\nu b_{\mu\nu} )  =0, \\
\partial^\mu \partial^\nu b_{\mu\nu} =0.
\end{eqnarray}

It is not difficult to check that the residual gauge
transformations (\ref{restr}) are sufficient to impose the gauge
\cite{Boos:2002ik} $$
\partial^\nu{ b_{\mu\nu}}=0,
$$ in which the former equations are satisfied identically.

Thus, in what follows we will be working in the gauge
\begin{eqnarray}\nonumber
(e^{-2A}h_{44})'-\frac{1}{3 M^3} e^{-2A}\phi'f=0, \\ \nonumber
h_{\mu 4} =0,\\ \label{compl_gauge} \tilde b = \gamma^{\mu\nu}{
b_{\mu\nu}}=0, \quad
\partial^\nu{ b_{\mu\nu}}=0,
\end{eqnarray}
 the residual gauge transformations now being
\begin{equation}\label{ok1}
\xi_\mu=e^{-2A}\epsilon_\mu(x),\qquad
\partial^\nu\epsilon_\nu(x)=0,\qquad \Box{\epsilon_\nu}=0.
\end{equation}

Obviously, after the substitution  $\tilde h=-2 h_{44}$ contracted
$\mu\nu$-equation  (\ref{cont}) and the $\mu4$-equation are
satisfied identically in this gauge. Eq. (\ref{munu}) for the
$\mu\nu$-component reduces to  an equation for a
transverse-traceless tensor field $b_{\mu\nu}(x,y)$:
\begin{equation}\label{ub}
\frac{1}{2}\left(e^{2A(y)}\Box
{b_{\mu\nu}}+\frac{\partial^2{b_{\mu\nu}}}{\partial
y^2}\right)-b_{\mu\nu}\left(2(A')^2-A''\right)=0.
\end{equation}
This equation does not include the background scalar filed
$\phi(y)$ and is absolutely analogous to the corresponding
equation in the unstabilized RS1 model.

In order to find equations for the scalar field $h_{44}$, we have
to solve gauge condition  (\ref{compl_gauge}) with respect to  $f$
and to substitute the latter into Eq. (\ref{44eq}), (\ref{eqf}).
This can be done either using the regularization $(sign(y))^2=1$,
or restricting the equations to the interval $(0,L)$ and taking
into account their singular terms with the help of boundary
conditions. The latter technique turns out to be simpler and we
will use it.

Equation for  44-component (\ref{44eq}) simplifies considerably,
when rewritten in the interval  $(0,L)$ in terms of a new function
$g =e^{-2A(y)}h_{44}(x,y)$ and using the expression for  the
potential  $V$ in terms of $A$ and $\phi$ (\ref{yd}):
\begin{equation}\label{ug0}
g'' +2g'\left(A'-\frac{\phi''}{\phi'}\right)-\frac{
(\phi')^2}{6M^3} g+
\partial_\mu \partial^\mu g =0.
\end{equation}
Let us note that substitution (\ref{hb}) and gauge condition
(\ref{compl_gauge}), which decouple the equations of motion, in
terms of $g$ take the form:
\begin{eqnarray}\label{subst}
h_{\mu\nu} = b_{\mu\nu} - \frac{1}{2} \eta_{\mu\nu} g, \quad
h_{44} = e^{2A(y)} g,
\\ \label{compl_gauge1}
g'-\frac{1}{3M^3}e^{-2A}\phi'f=0,
\\ \label{compl_gauge2}
 h_{\mu 4} =0, \quad  \tilde b = \gamma_{\mu\nu} b^{\mu\nu} =0, \quad
\partial^\nu{b_{\mu\nu}}=0.
\end{eqnarray}

Gauge condition  (\ref{compl_gauge1}) solved for  $f$ in the
interval $(0,L)$, looks like $$ f=3M^3\frac{e^{2A}}{\phi'}g'. $$
Substituting this expression for  $f$ into Eq. (\ref{eqf}) gives
an equation, which is obtained by differentiating Eq. (\ref{ug0})
with respect to  $y$,   and boundary conditions on the branes:
\begin{eqnarray}\nonumber
\left(\frac{1}{2}\frac{d^2 \lambda_1}{d \phi^2} -
\frac{\phi''}{\phi'}\right)g' +\partial_\mu
\partial^\mu g|_{y=+0}=0, \\ \label{bc}
\left(\frac{1}{2}\frac{d^2\lambda_2}{d \phi^2} +
\frac{\phi''}{\phi'}\right)g' -\partial_\mu \partial^\mu
g|_{y=L-0}=0.
\end{eqnarray}
Thus, our gauge choice and the substitution enabled us to decouple
the equations of motion.  This means that the fluctuations of the
metric and of the scalar  field against any background given by a
solution to Eqs. (\ref{yd}) are described by two fields, -- tensor
field  $b_{\mu\nu}(x,y)$ and  scalar field $g(x,y)$. Their
classical equations of motion are given by Eqs. (\ref{ub}) and
(\ref{ug0}), (\ref{bc}) respectively.

\section{Mode decompositions}
Let us study first the modes of the tensor field
$b_{\mu\nu}(x,y)$, which satisfies Eq. (\ref{ub}). Substituting
into this equation $$ b_{\mu\nu}(x,y) = c_{\mu\nu}e^{ipx}
\psi_n(y),\quad c_{\mu\nu} = const, \quad p^2 = -m_n^2, $$
restricting it to the interval $(0,L)$ and replacing the singular
terms by the boundary conditions, we get:
\begin{eqnarray}\nonumber
\frac{d^2 \psi_n}{dy^2} -2(2(A')^2 -A'')\psi_n = -m_n^2 e^{2A}
\psi_n, \\ \label{bmode}  \psi_n^\prime + 2 A'  \psi_n   |_{y=+0}=
\psi_n^\prime + 2 A'  \psi_n |_{y=L-0}=0.
\end{eqnarray}
The boundary conditions suggest a substitution $\psi_n =
\exp(-2A)\omega_n $, which turns this equation into
\begin{eqnarray}\nonumber
\frac{d}{dy} \left(e^{-4A}\omega_n'\right) = -m_n^2 e^{-2A}
\omega_n, \\ \label{bmode1}  \omega_n' |_{y=+0}=
\omega_n'|_{y=L-0}=0.
\end{eqnarray}
We see that the eigenfunctions $\omega_n$ are solutions of a
Sturm-Liouville problem with von Neumann boundary conditions. In
accordance with the general theory,\cite{BKM} the problem at hand
has no negative eigenvalues for arbitrary $A$, only one zero
eigenvalue, corresponding to $\omega_0 = const$, and an infinite
number of positive eigenvalues, asymptotically given by the
formula
\begin{equation}\label{bmass}
m_n^2 = \frac{\pi^2 n^2}{l^2}, \quad l = \int_0^L e^{A(y)} dy.
\end{equation}
This formula should be specified for finding the masses on
different branes. We recall that the masses of excitations on each
brane should be calculated in the Galilean coordinates
\cite{Rubakov:2001kp,Boos:2004uc,Boos:2002ik}, for which $A(y)$ is
equal to zero on the corresponding brane (we recall that
coordinates are called Galilean if $g_{\mu\nu}= (-1,1,1,1)$
\cite{LL}). Thus, the latter formula can be explicitly adopted for
calculating masses on the branes as follows:
\begin{eqnarray}\label{asymp0}
&l = \int_0^L e^{(A(y)- A(0))} dy \quad \makebox{for the brane
at}\quad y=0,&\\ \label{asympL} &l = \int_0^L e^{(A(y)-A(L))} dy
\quad \makebox{for the brane at}\quad y=L,&
\end{eqnarray}
which is valid for an arbitrary $A(y)$ satisfying  Eqs.
(\ref{yd}), because they define it  up to an additive constant.

The eigenfunctions $\{\psi_n(y)\}$ of eigenvalue problem
(\ref{bmode}) build a complete orthonormal set, the eigenfunction
of the zero mode being
\begin{equation} \label{zeromode}
\psi_0(y) = N e^{-2A(y)}.
\end{equation}
 Expanding  $b_{\mu\nu}$ in this system
\begin{equation}\label{decomp}
b_{\mu\nu}=\sum_{n=0}^\infty b_{\mu\nu}^n(x)\psi_n(y),
\end{equation}
we get four-dimensional tensor fields $b_{\mu\nu}^n(x)$ with
definite masses. An important point is that due to the form of the
zero mode eigenfunction residual gauge transformations (\ref{ok1})
act only on the massless field $b_{\mu\nu}^0(x)$ and provide the
correct number of  degrees of freedom  of the massless graviton
\cite{Boos:2002ik}.

In order to find the mass spectrum of the scalar particles
described by Eq. (\ref{ug0}) let us substitute $$ g(x,y) = e^{ipx}
g_n(y), \quad p^2 = -\mu_n^2, $$ into this equation. As a result,
the equation and the boundary conditions for $ g_n(y)$ take the
form:
\begin{eqnarray}\label{eq_bc}
g_n'' +2A'g_n'-2\frac{\phi''}{\phi'}g_n'-\frac{ (\phi')^2}{6M^3}
g_n= - \mu_n^2 e^{2A}g_n, \\ \label{eq_bc1}
\left(\frac{1}{2}\frac{d^2 \lambda_1}{d \phi^2} -
\frac{\phi''}{\phi'}\right)g_n' + \mu_n^2e^{2A}g_n|_{y=+0}=0, \\
\label{eq_bc2} \left(\frac{1}{2}\frac{d^2\lambda_2}{d \phi^2} +
\frac{\phi''}{\phi'}\right)g_n'  - \mu_n^2e^{2A}g_n|_{y=L-0}=0.
\end{eqnarray}

 Let us write Eq. (\ref{eq_bc}) in the Sturm-Liouville form:
\begin{equation}\label{eq_sc}
\frac{d}{dy}\left(\frac{e^{2A}}{(\phi')^2} g_n'\right)
-\frac{e^{2A}}{6M^3}g_n=-\mu_n^2 g_n\frac{e^{4A}}{(\phi')^2}.
\end{equation}
It is not difficult to see that the operator in the eigenvalue
problem for this equation with boundary conditions (\ref{eq_bc1}),
(\ref{eq_bc2}) is not self-adjoint. Nevertheless, these boundary
conditions look like the usual Sturm-Liouville boundary conditions
and lead to a number of general assertions about the spectrum and
the eigenfunctions of the problem at hand.

Multiplying Eq. (\ref{eq_sc}) by $\bar g_n$, then integrating over
$(0,L)$ and integrating by parts in the term with derivatives, we
get:
\begin{eqnarray}\nonumber
\mu_n^2 \left( \int_0^L \frac{e^{4A}}{(\phi')^2}|g_n|^2 dy +
\left(\frac{1}{2}\frac{d^2 \lambda_1}{d \phi^2} -
\frac{\phi''}{\phi'}\right)^{-1}
 \frac{e^{4A}}{(\phi')^2}|g_n|^2 |_{y=+0}+ \right. \\ \nonumber
 +\left.
\left(\frac{1}{2}\frac{d^2 \lambda_2}{d \phi^2} +
\frac{\phi''}{\phi'}\right)^{-1} \frac{e^{4A}}{ (\phi')^2}|g_n|^2
|_{y=L-0}\right) =  \\ \nonumber =\frac{1}{6M^3}  \int_0^L
e^{2A}|g_n|^2 dy +  \int_0^L \frac{e^{2A}}{(\phi')^2} |g_n'|^2 dy.
\end{eqnarray}
This means that if
\begin{equation}\label{parconstr}
\left(\frac{1}{2}\frac{d^2 \lambda_1}{d \phi^2} -
\frac{\phi''}{\phi'}\right)|_{y=+0}>0, \quad
\left(\frac{1}{2}\frac{d^2\lambda_2}{d \phi^2} +
\frac{\phi''}{\phi'}\right)|_{y=L-0}>0,
\end{equation}
all the eigenvalues of the eigenproblem are real and positive. The
standard technique for proving the orthogonality of eigenfunctions
gives in this case for  $m\neq n$
\begin{eqnarray}\nonumber
- \int_0^L \frac{e^{4A}}{(\phi')^2}\bar g_m g_n dy
=\left(\frac{1}{2}\frac{d^2 \lambda_1}{d \phi^2} -
\frac{\phi''}{\phi'}\right)^{-1} \frac{e^{4A}}{(\phi')^2}\bar g_m
g_n |_{y=+0}+ \\ \label{eq_ef}  + \left(\frac{1}{2}\frac{d^2
\lambda_2}{d \phi^2} + \frac{\phi''}{\phi'}\right)^{-1}
\frac{e^{4A}}{ (\phi')^2}\bar g_m g_n |_{y=L-0}.
\end{eqnarray}
Thus, the eigenfunctions of this problem, corresponding to
different eigenvalues, are not orthogonal with respect to the
weight suggested by the form of Eq. (\ref{eq_sc}).

It is very difficult to prove rigorously that this set of
eigenfunctions is complete.  We can just argue that due to the
special form of boundary conditions (\ref{eq_bc1}), (\ref{eq_bc2})
the set of eigenfunctions of the eigenvalue problem under
consideration is complete. Really, for the parameters of the
scalar field potential  $
\frac{1}{2}\frac{d^2\lambda_{1,2}}{d\phi^2} \to \infty$ the
eigenvalues drop from the boundary conditions, and the operator in
the equation becomes self-adjoint (this approximation was used in
\cite{Csaki:2000zn}, \cite{Kofman:2004tk}). Therefore, for
$\frac{1}{2}\frac{d^2\lambda_{1,2}}{d\phi^2} \to \infty$ the
eigenfunctions of the problem under consideration go to the
eigenfunctions of a Sturm-Liouville problem with a self-adjoint
operator, which build a complete orthogonal set. The orthogonality
of the eigenfunctions for
$\frac{1}{2}\frac{d^2\lambda_{1,2}}{d\phi^2} \to \infty$ can be
also seen in Eq. (\ref{eq_ef}).  Thus, we assume that the
eigenfunctions $\{g_n(y)\}$  of the problem  (\ref{eq_bc}) -
(\ref{eq_bc2}) form a complete denumerable  non-orthogonal set. We
would also like to note that for  $ \mu_n^2\to \infty$ boundary
conditions  (\ref{eq_bc1}), (\ref{eq_bc2})  go to $g_n(0) =g_n(L)
= 0$, and therefore for large  $m$ or $n$ the integral in
(\ref{eq_ef}) is close to zero. Such a set of functions may be
called asymptotically orthogonal.

It is easy to understand that the eigenfunctions $g_n(y)$ can be
chosen to be real. Then five-dimensional scalar field $g(x,y)$ can
be expanded in this system as
\begin{equation}\label{decomps}
g(x,y) =\sum_{n=1}^\infty \varphi_n(x)g_n(y),
\end{equation}
where the four dimensional real scalar fields $\varphi_n(x)$ have
masses $\mu_n^2$.

Now we can find the effective four-dimensional action for the
system. Substituting  (\ref{hb}),
(\ref{subst})-(\ref{compl_gauge2}) into Lagrangian (\ref{l}),  we
find that the tensor field and the scalar field Lagrangians
decouple. Then substituting  expansions (\ref{decomp}),
(\ref{decomps}) into these Lagrangians  and using Eqs.
(\ref{bmode}), (\ref{eq_bc}), assuming the eigenfunctions
$\{\psi_n\}$ and $\{g_n\}$ to be appropriately normalized and
integrating over the coordinate of the extra dimension, we find
that the reduced tensor field Lagrangian is a sum of the standard
four-dimensional Lagrangians of tensor fields with masses $m_n$,
whereas for the scalar field Lagrangian we get
\begin{eqnarray}\nonumber
{\mathcal{L}}_{scalar} =-\frac{3}{4}\sum_{nk}
\left[\eta^{\mu\nu}\partial_\mu
\varphi_n\partial_\nu\varphi_k+\mu_k^2 \varphi_n \varphi_k
\right]\int_{0}^{L}
dye^{2A}\left(g_{n}g_{k}+\frac{6M^{3}}{(\phi')^2}g'_{n}g'_{k}\right).
\end{eqnarray}
Now consider the integral over $dy$. Integrating by parts  the
term with derivatives, using Eq. (\ref{eq_sc}) and boundary
conditions (\ref{eq_bc1}), (\ref{eq_bc2}), we arrive at the result
\begin{eqnarray*}
6\mu_n^2 M^3\left( \int_0^L \frac{e^{4A}}{(\phi')^2}g_n g_k dy +
\left(\frac{1}{2}\frac{d^2 \lambda_1}{d \phi^2} -
\frac{\phi''}{\phi'}\right)^{-1}
 \frac{e^{4A}}{(\phi')^2}g_n g_k |_{y=+0}+ \right. \\
+\left. \left(\frac{1}{2}\frac{d^2 \lambda_2}{d \phi^2} +
\frac{\phi''}{\phi'}\right)^{-1} \frac{e^{4A}}{ (\phi')^2}g_n g_k
|_{y=L-0}\right).
\end{eqnarray*}
For real  $g_n(y)$ Eq. (\ref{eq_ef})  now  implies that this
expression is equal to zero for $n\neq k$. Assuming the
eigenfunctions $\{g_n\}$ to be appropriately normalized, we find
that the reduced scalar field Lagrangian is also the standard
Lagrangian, and the complete reduced action is
\begin{eqnarray}\nonumber
S_{eff}=-\frac{1}{4}\sum_{k}\int dx \left(
\partial^\sigma b^{k,\mu \nu}\partial_\sigma b_{\mu \nu}^k+m_k^2
b^{k,\mu\nu}b_{\mu \nu}^k \right) -\\ \label{redact}
-\frac{1}{2}\sum_{k}\int dx \left(\partial_\nu
\varphi_k\partial^\nu \varphi_k+\mu_k^2\varphi_k \varphi_k
\right).
\end{eqnarray}
We see that the effective Lagrangian of the tensor fields
coincides with the one of the unstabilized
model,\cite{Boos:2002ik} whereas the  effective Lagrangian of the
scalar fields coincides with the one found in \cite{Kofman:2004tk}
for the case $\frac{d^2 \lambda_{1,2}}{d \phi^2}\to\infty$.

Now  we are able to find the couplings  of the four-dimensional
fields $b_{\mu\nu}^n(x)$ and $\varphi_n(x)$, to matter on the
branes, which are defined by the coupling  of the fluctuations of
the five-dimensional gravitational field   $h_{\mu\nu}$ to matter
on the branes.  The latter is given by
\begin{eqnarray}\nonumber
\frac{1}{\sqrt{8M^3}}\int_{B_1}h^{\mu\nu}(x,0)T^{(1)}_{\mu\nu}\sqrt{-de
t\gamma_{\mu\nu}(0)}dx+ \\ \label{vz} +\frac{1}{\sqrt{8M^3}}
\int_{B_2}h^{\mu\nu}(x,L)T^{(2)}_{\mu\nu}\sqrt{-det\gamma_{\mu\nu}(L)}dx,
\end{eqnarray}
where  $T^{(1)}_{\mu\nu}$ and $T^{(2)}_{\mu\nu}$ are
energy-momentum tensors on brane 1 and 2, respectively and
\begin{eqnarray}\nonumber
T^{(1,2)}_{\mu\nu}=-\left[2\frac{\partial L^{(1,2)}}{\partial
g^{\mu\nu}}-g_{\mu\nu}L^{(1,2)}\right]
\end{eqnarray}
for our choice of the metric signature.

In what follows we restrict ourselves to considering the fields on
the brane at  $y=L$ only, which we  assume to be "our"\ brane.
Substituting decompositions (\ref{decomp}), (\ref{decomps}) into
(\ref{vz}) we find that the interaction of the tensor and the
scalar fields with matter on the brane at  $y=L$ in the Galilean
coordinates is
\begin{eqnarray}\nonumber
\frac{1}{\sqrt{8M^3}}\int_{B_2}\left(\psi_0(L)b_
{\mu\nu}^0(x)T^{\mu\nu}+\sum_{n=1}^\infty\psi_n(L)b_{\mu\nu}^n(x)T^{\mu\nu}-\right.
\\ \label{allint} - \left.\frac{1}{2}\sum_{n=1}^\infty  g_n(L) \varphi_n(x) T_\mu^\mu
\right)dx.
\end{eqnarray}
Thus,  the couplings are defined by the values of the wave
functions on the brane. The latter can be found only if we specify
the background solutions $A(y)$ and $\phi(y)$.

\section{Specific example}
The considerations in the previous sections allowed us to find the
general structure of the  brane world models stabilized by the
scalar field. To make  predictions for the masses and the coupling
constants we must take specific potentials for the scalar field
and a particular  vacuum solution of system (\ref{yd}).

To find an analytic solution to this system we will use the
results of \cite{wolfe}, \cite{Brandhuber}. Let us consider a
special class of potentials, which can be represented as $$
V(\phi)=\frac{1}{8} \left(\frac{d W}{d\phi}\right)^2-
\frac{1}{24M^3}W^2(\phi). $$ It is easy to check that if we put
\begin{equation}
\phi'(y) =sign(y)\frac{1}{2} \frac{d W}{d\phi}, \quad A'(y) =
sign(y)\frac{1}{24M^3}W(\phi),
\end{equation}
then Eqs. (\ref{yd}) are valid everywhere, except for the branes.
In order the equations of motion be valid everywhere, one needs to
finetune the brane potentials $\lambda_{1,2}(\phi)$.

Let us take  $W(\phi)$ to be
\begin{equation}
W=24M^3 k - u \phi^2,
\end{equation}
so that $V(\phi)$ is a quartic potential. Finetuned potentials on
the branes can be chosen as follows:
\begin{eqnarray}
\lambda_{1}= W(\phi_1)+
W'(\phi_1)(\phi-\phi_1)+\beta^2_1(\phi-\phi_1)^2, \\ \lambda_{2}=-
W(\phi_2)- W'(\phi_2)(\phi-\phi_2)+\beta^2_2 (\phi-\phi_2)^2.
\end{eqnarray}
The parameters of the potentials  $k, u, \phi_{1,2}, \beta_{1,2}$,
when made dimensionless by the fundamental five-dimensional energy
scale of the theory $M$, should be positive quantities of the
order $O(1)$, i.e. there should be no hierarchical difference in
the parameters.

For such a choice of the potentials the solution of the equations
of motion is given by  \cite{wolfe}
\begin{eqnarray}
\phi(y)=\phi_1\, e^{-u|y|}\\ A(y)=k|y|+\frac{
\phi^2_1}{48M^3}e^{-2u|y|}.
\end{eqnarray}
The interbrane distance is defined by the boundary conditions for
the field $\phi$  and is expressed in terms of the parameters of
the model by the relation
\begin{equation}
L= \frac{1}{u} \ln\left(\frac{\phi_1}{\phi_2}\right).
\end{equation}
Thus, we see that the brane separation distance is stabilized.

Let us study the mass spectrum of tensor particles, which is
defined by Eq. (\ref{bmode}). The zero mode solution for arbitrary
$A$ is given by (\ref{zeromode}). For our choice of $A$ it is
impossible to find exact solutions for other modes. Therefore, in
what follows we will use an approximation $uL\ll1$, which is
rather general and physically interesting \cite{Boos:2004uc}.
Keeping in  $A$ only the terms linear in $y$, we get
\begin{equation}\label{renormk}
A(y)= \tilde{k}|y|, \quad \tilde k = k-\frac{\phi_1^2}{24M^3}u.
\end{equation}
Thus, in this approximation Eq. (\ref{ub}) for the tensor field
coincides with the equation of the unstabilized model, where a
substitution
 $k\to \tilde k$ was made. This equation can be solved exactly, and the formulas
for eigenfunctions and eigenvalues were discussed in detail in
\cite{Boos:2002ik}. In particular, in this approximation the
normalized functions $\psi_0$ for  $m_0=0$ look like
\begin{equation}\label{57}
\psi_0(y)=N_0 e^{-2\tilde{k}|y|},\qquad
N_0=\frac{\tilde{k}^{\frac{1}{2}}}{(1-e^{-2\tilde{k}L})^\frac{1}{2}}
\end{equation}
in the coordinates, which are Galilean on the brane at $y=0$, and
\begin{equation}\label{59}
\psi_0(y)=N_0 e^{-2\tilde{k}|y|+ 2\tilde{k}L},\qquad
N_0=\frac{\tilde{k}^{\frac{1}{2}}}
{(e^{2\tilde{k}L}-1)^\frac{1}{2}}
\end{equation}
in the coordinates, which are Galilean on the brane at $y=L$. For
the first massive tensor excitation on the brane at $y=L$ we
get\cite{Boos:2002ik} $\psi_1(L)\approx -\sqrt{\tilde k}$, $ m_1
\approx 3.83\tilde k$. Eqs. (\ref{bmass}), (\ref{asymp0}),
(\ref{asympL}) give the following mass spectra for large $n$:
\begin{eqnarray}\nonumber
m_n^2 = {\pi^2 {\tilde k}^2 n^2} e^{-2{\tilde k}L} \quad
(\makebox{the brane at}\quad y=0), \\ \label{masstall}  m_n^2 =
{\pi^2 {\tilde k}^2 n^2} \quad (\makebox{the brane at}\quad y=L).
\end{eqnarray}

Let us study now the mass spectra of the scalar particles.
Equation and boundary conditions (\ref{eq_bc}) - (\ref{eq_bc2})
for our choice of the potentials and background solutions take the
form:
\begin{equation}\label{eq_bcE}
g_n'' +2A'g_n'-2\frac{\phi''}{\phi'}g_n'-\frac{ (\phi')^2}{6M^3}
g_n= - \mu_n^2 e^{2A}g_n,
\end{equation}
\begin{equation}\label{eq_bcE1}
(\beta_1^2 +u)g_n' + \mu_n^2e^{2A}g_n|_{y=+0}=0,
\end{equation}
\begin{equation}\label{eq_bcE2}
(\beta_2^2 -u)g_n' - \mu_n^2e^{2A}g_n|_{y=L-0}=0.
\end{equation}

 To find explicitly the eigenfunctions and eigenvalues of the problem we will
use the same approximation  $uL \ll 1$,   which now implies $$ uy
= uL \frac{y}{L} < uL \ll 1. $$ Substituting the explicit form of
$\phi$ into  (\ref{eq_bcE}) we get
\begin{equation}\label{ug3}
g_n'' +2A'g_n'+2ug_n'-\frac{ \phi_1^2}{6M^3} u^2 g_n+\mu_n^2
e^{2A}g_n =0.
\end{equation}

We solve this equation in the coordinates, which are Galilean on
the brane at $y=L$. In this case  $A(y)$ is expressed in terms of
$\tilde k$ (\ref{renormk}) as: $$ A(y) = \tilde k (y -  L). $$ We
have already found that for $\beta_2^2 -u>0$ all eigenvalues of
this problem are larger than zero. Therefore, we introduce a new
variable by the relation $$ z=\frac{\mu_n}{\tilde k} e^{\tilde k y
- \tilde k L}, \quad \frac{\mu_n}{\tilde k} e^{ - \tilde k L} \leq
z \leq \frac{\mu_n}{\tilde k} . $$ In terms of this variable the
equation takes the form:
\begin{equation}\label{ug7}
\frac{d^2 g_n}{dz^2} +\left(3+ \frac{2u}{\tilde k} \right)
\frac{1}{z}\frac{dg_n}{dz } + \left(1- \frac{b^2}{z^2} \right) g_n
=0, \quad b^2 = \frac{ \phi_1^2}{6M^3} \frac{u^2}{{\tilde k}^2}.
\end{equation}
Let us look for $g_n$ in the form $g_n(z)= z^a t_n(z)$. Then the
equation for $t_n$ is
\begin{equation}\label{ug4}
\frac{d^2t_n}{dz^2 } +\left(2a +3+ \frac{2u}{\tilde k} \right)
\frac{1}{z}\frac{dt_n}{dz } + \frac{a\left(a +2+ \frac{2u}{\tilde
k} \right)}{{z}^2}t_n + \left(1- \frac{b^2}{z^2} \right) t_n =0.
\end{equation}
In order to turn this equation into the Bessel equation we put $a
=-\left(1+ \frac{u}{\tilde k} \right)$ and get
\begin{equation}\label{ug5}
\frac{d^2t_n}{dz^2 } + \frac{1}{z}\frac{dt_n}{dz } + \left(1-
\frac{\alpha^2}{z^2} \right) t_n =0, \quad \alpha^2 = a^2 + b^2.
\end{equation}
The general solution to this equation is $$ t_n(z)= AJ_\alpha(z) +
B J_{-\alpha}(z), \quad \alpha = \sqrt{a^2 + b^2}. $$
Correspondingly, we get
\begin{equation}\label{solg}
g_n (z)= z^{-\left(1+ \frac{u}{\tilde k}
\right)}\left(AJ_\alpha(z) + B J_{-\alpha}(z)\right).
\end{equation}
The boundary conditions in terms of  $z$ look like:
\begin{eqnarray}\nonumber
{\tilde k z^{2}} g_n+(\beta^2_1+u)z\frac{d g_n}{dz}|_{z_1 =\frac{
\mu_n}{\tilde k} e^{ - \tilde k L}} =0, \\ {\tilde k z^{2}}
g_n-(\beta^2_2-u)z\frac{d g_n}{dz}|_{z_2 =\frac{\mu_n}{\tilde k}}
=0.
\end{eqnarray}

Below we show that for reproducing the Newtonian gravity on the
brane at $y=L$ for strong five-dimensional gravity we must take
$\tilde kL \sim 35$. In this case   ${z_1 =\frac{\mu_n}{\tilde k}
e^{- \tilde k L}}\approx 0$ is a very good approximation, and the
boundary condition at zero allows us to drop the singular term
with $J_{-\alpha}(z)$ in  $g_n(z)$ since  $B/A \sim e^{ - 2\tilde
k L},$ and the corrections due to this term are negligible. Thus,
up to normalization  $g_n(z)$ can be written as $$ g_n(z) =
z^{-\left(1+ \frac{u}{\tilde k} \right)}J_\alpha(z).$$ The second
boundary condition at  ${z_2=\frac{\mu_n}{\tilde k}}$ gives an
equation for the mass spectrum of the scalar particles:
\begin{equation}\label{spectrum}
\left(1+ \alpha + \frac{u}{\tilde k}+ \frac{\tilde k
z_2^2}{\beta_2^2 -u }\right)J_{\alpha}(z_2) - z_2
J_{\alpha-1}(z_2)  =0.
\end{equation}
Expanding the Bessel function for small $z_2$ up to the second
term and keeping the terms up to the order  $z_2^2$ in the
equation, we get the following relation for the mass of the lowest
scalar excitation:
\begin{equation}\label{mradion}
\mu_1^2 = \frac{4 \tilde k^2 (-1+\alpha - \frac{u}{\tilde k})(1+
\alpha)(\beta_2^2 -u)}{4\tilde k (1+ \alpha)+ (1+ \alpha -
\frac{u}{\tilde k})(\beta_2^2 -u)}.
\end{equation}
For small  $u/\tilde k$ it reduces to
\begin{equation}\label{mrad}
\mu_1^2 = \frac{\phi_1^2 u^2}{3M^3}\, \frac{\beta_2^2
-u}{\beta_2^2 +4\tilde k},
\end{equation}
which for  $\beta_2^2 \to \infty$ formally coincides with the
results of \cite{Csaki:2000zn}, \cite{Kofman:2004tk},
\cite{Boos:2004uc} in our approximation (we recall that the
fundamental energy scale in \cite{Csaki:2000zn},
\cite{Kofman:2004tk} is the Planck one). The next roots of Eq.
(\ref{spectrum}) are of the order $\tilde k^2$, and the asymptotic
formula of the general theory \cite{BKM} for large $n$ gives
$\mu_n^2 = \pi^2 \tilde k^2 n^2$.

The normalization condition for the eigenfunctions $\{g_n(y),
n=0,1,...\}$ of the problem (\ref{eq_bc}) - (\ref{eq_bc2}), which
gives the canonical kinetic terms in  (\ref{redact}), is the
following:
\begin{equation}\label{norm}
\frac{3}{2}\int_{0}^{L} dy
e^{2A}\left(g_{n}g_{k}+\frac{6M^3}{(\phi')^2}g'_{n}g'_{k}\right)=
\delta_{nk}.
\end{equation}

The normalized functions  $g_n(y)$ look like
\begin{equation}\label{solg1}g_n (y)=A_n
\left(\frac{\mu_n}{\tilde k} e^{\tilde k y - \tilde
kL}\right)^{-\left(1+ \frac{u}{\tilde k}
\right)}J_\alpha\left(\frac{\mu_n}{\tilde k} e^{\tilde k y -
\tilde kL}\right),
\end{equation}

\begin{eqnarray}\nonumber
 A_n &=& \frac{ u \phi_1\left(\mu_n/\tilde k\right)^{1+u/\tilde k}e^{-uL}}{3M^{\frac{3}{2}} \mu_n J_{\alpha}\left({\mu_n}/{\tilde k}\right)}\times
 \\ \nonumber & \times &\left[\frac{1}{2\tilde k}\left[\left(\frac{\tilde k +u}{\mu_{n}}+\frac{\mu_{n}}{\beta_{2}^{2}-u}\right)^{2}+
 \left(1-\frac{{\tilde k}^{2}\alpha^{2}}{\mu_{n}^{2}}\right)\right]+\frac{1}{\beta_{2}^{2}-u}\right]^{-\frac{1}{2}}\,\,.
\end{eqnarray}
We also note that in our approximation the boundary term at $y=0$
drops from the normalization condition.

Now that we have found explicit expressions for the wave functions
of the tensor and the scalar fields, we can find their coupling
constants to matter on the brane at $y=L$.

The coefficient in front of the zero mode of the tensor field
$b_{\mu\nu}^0(x)$ in (\ref{vz}) can be expressed in terms of the
Planck mass on the brane at  $y=L$ as $1/\sqrt{8 M_{Pl}^2}$.  Then
Eq. (\ref{59}) implies:
\begin{equation}
\frac{1}{\sqrt{{8 M_{Pl}^2}}}=\frac{1}{\sqrt{8M^3}}N_0
=\frac{1}{\sqrt{8M^3}}\frac{\tilde{k}^{\frac{1}{2}}}
{(e^{2\tilde{k}L}-1)^\frac{1}{2}},
\end{equation}
which gives a relation between  the Planck mass on the negative
tension brane and the fundamental five-dimensional energy scale
$M$
\begin{equation}
M_{Pl}^2=M^3 \frac{e^{2\tilde{k}L}-1}{\tilde{k}}.
\end{equation}

Again we have obtained the formulas, which coincide with the
formulas of the unstabilized model after replacing $k \to \tilde
k$. Therefore, in order to get the correct value of the
gravitational constant on the brane at $y=L$ one can take the same
values of the parameters, which are used in the unstabilized
model, i.e. $M \sim \tilde k \sim 1 TeV, \quad \tilde k L \sim 35$
\cite{Boos:2004uc}. Then the parameter  $u$ may be of the order of
ten $GeV$,  and the radion mass  (\ref{mrad}) may be of the order
of hundred  $GeV$.

The exact solution for the zero mode of the tensor field
(\ref{zeromode}) allows one to get the relation between the Planck
mass on the brane at  $y=L$ and the fundamental five-dimensional
energy scale beyond the approximation used here. Other possible
approximations were studied in \cite{Boos:2004uc}.

Eq. (\ref{allint}) shows  that the coupling constant of the n-th
scalar mode to matter is defined by the value of its wave function
on the brane at $y=L$ and is approximately given by the relation
$$ \epsilon_n=-\frac{g_{n}(L)}{2\sqrt{8M^{3}}}=
-\frac{A_n}{\sqrt{32M^3}} J_\alpha \left(\frac{\mu_n}{\tilde
k}\right) \left(\frac{\mu_n}{\tilde k}\right)^{-(1+u/\tilde k)}.
$$ For the radion, this constant is $$ \epsilon_1\simeq
-\sqrt{\frac{\tilde k}{24M^3}},$$ and turns out to be of the order
$\epsilon_1^{-1}\sim 5 TeV$ for the above given values of the
model parameters. One can see that $\epsilon_1$ does not depend on
parameter $\beta_{2}$.

\section{Discussion}
In the present paper we have considered the general structure of
the brane world models stabilized by the scalar field. For an
arbitrary background configuration of the gravitational and scalar
fields, satisfying the equations of motion for a stabilized model,
we constructed the second variation Lagrangian and derived the
equations for the fields describing the fluctuations against the
background. A convenient gauge and a substitution were found,
which enabled us to decouple the equations of motions and to
isolate the five-dimensional degrees of freedom. It was shown that
the tensor sector splits from the scalar one and has the same
structure, as in the unstabilized model. Namely, for any
background there is a massless four-dimensional graviton and a
tower of massive tensor fields, which represent the
four-dimensional degrees of freedom in the tensor sector.

The structure of the scalar sector was found to be more
complicated: the operator of the mass squared turned out to be
non-self-adjoint and its eigenfunctions to be non-orthogonal.
Nevertheless, the structure of the five-dimensional action is such
that the non-diagonal interaction terms of the four-dimensional
scalar modes vanish.

For a particular choice of the background solution and the
parameters of the model it was found that the influence of the
scalar field background on the tensor excitations reduces to a
renormalization of the parameter $k$ of the unstabilized model,
which is replaced by $\tilde k$  (\ref{renormk}). In this case the
inverse size of extra dimension may be of the order of tens of GeV
and the masses of tensor excitations may be of the order of a few
$TeV$, the radion mass being of the order of $100 GeV$. For this
choice of the parameters the radion coupling constant to matter on
the brane at $y=L$ turned out to be of the same order as in the
unstabilized model with the same choice of the parameters.

At the same time it is quite possible that it can be much larger
for a different choice of the model parameters. In this case the
radion mass and the masses of the tensor excitations must also
shift. To find out, whether it is really so,  one has to scan the
whole parameter space of the model. The coupling of the radion to
matter can also become stronger due to the radion-Higgs mixing as
discussed, for example, in \cite{Csaki:2000zn},
\cite{Giudice:2000av}--\cite{Hewett:2002nk}.

\section*{Acknowledgments}

The work of E. B., M. S. and I. V. is partly supported by
RFBR~04-02-17448, Universities of Russia UR.02.02.503, and Russian
Ministry of Education and Science NS-8122.2006.2 grants. M.S. also
acknowledges support of grant for young scientists MK-8718.2006.2
of the President of Russian Federation. The authors are grateful
to Yu.~Grats, P.~Ermolov, Yu.~Kubyshin and V.~Rubakov for useful
discussions. I.V. would also like to thank A.~Belyaev,
S.~Chivukula and C.-P.~Yuan for interesting discussions.

\end{document}